\documentclass{emulateapj}

\usepackage{rotating}
\newcommand{\mbh}{$M_{\rm BH}$}
\newcommand{\mss}{$M_{\rm sph, \star}$}
\newcommand{\mts}{$M_{\rm host, \star}$}
\newcommand{\mstar}{$M_{\rm \star}$}

\begin{document}

\title{THE RELATION BETWEEN BLACK HOLE MASS AND HOST SPHEROID STELLAR MASS OUT TO $z\sim2$}

\shorttitle{CO-EVOLUTION OF SPHEROIDS AND BLACK HOLES.}
\shortauthors{Bennert et al.}

\author{Vardha N. Bennert\altaffilmark{1,2},
Matthew W. Auger\altaffilmark{1}, 
Tommaso Treu\altaffilmark{1,3}, 
Jong-Hak Woo\altaffilmark{4}, 
Matthew A. Malkan\altaffilmark{5}}

\altaffiltext{1}{Department of Physics, University of California,
Santa Barbara, CA 93106; tt@physics.ucsb.edu; mauger@ast.cam.ac.uk}

\altaffiltext{2}{Current address:
Physics Department, California Polytechnic State University,
San Luis Obispo, CA 93407; vbennert@calpoly.edu}

\altaffiltext{3}{Packard Fellow}

\altaffiltext{4}{Astronomy Program, Department of Physics and Astronomy, Seoul National University, 
Korea; woo@astro.snu.ac.kr}

\altaffiltext{5}{Department of Physics and Astronomy, University of California,
Los Angeles, CA 90095; malkan@astro.ucla.edu}

\shortauthors{Bennert et al.}

\begin{abstract}
We combine Hubble Space Telescope images from the Great Observatories
Origins Deep Survey with archival Very Large Telescope and Keck
spectra of a sample of 11 X-ray selected broad-line active galactic
nuclei in the redshift range $1<z<2$ to study the black hole mass -
stellar mass relation out to a lookback time of 10 Gyrs. Stellar
masses of the spheroidal component (\mss) are derived from
multi-filter surface photometry.  Black hole masses (\mbh) are
estimated from the width of the broad MgII emission line and the
3000\AA~nuclear luminosity. Comparing with a uniformly measured
local sample and taking into account selection effects,
we find evolution in the form
\mbh/\mss $\propto(1+z)^{1.96\pm0.55}$, in agreement with our earlier
studies based on spheroid luminosity.  However, this result is more
accurate because it does not require a correction for luminosity
evolution and therefore avoids the related and dominant systematic
uncertainty.  We also measure total stellar masses (\mts). Combining
our sample with data from the literature, we find \mbh/\mts $\propto$
$(1+z)^{1.15\pm0.15}$, consistent with the hypothesis that black
holes (in the range \mbh~$\sim10^{8-9}$M$_\odot$) predate the formation of
their host galaxies.  Roughly one third of our objects reside in
spiral galaxies; none of the host galaxies reveal
signs of interaction or major merger activity.
Combined with the slower evolution in host stellar masses 
compared to spheroid stellar masses, 
our results indicate that secular evolution or minor mergers
play a non-negligible role in growing both BHs and spheroids.
\end{abstract}


\section{INTRODUCTION}
\label{sec:intro}
Active Galactic Nuclei (AGNs) are thought to represent an integral
phase in the formation and evolution of galaxies during which the
central supermassive black hole (BH) is growing through accretion. The
empirical relations between BH mass (\mbh) and the properties of the
host galaxy \citep[e.g.,][]{fer00,geb00,mar03,har04} have been
explained by a combination of AGN feedback
\citep[e.g.,][]{vol03,cio07,dim08,hop09} and hierarchical assembly of
\mbh~and stellar mass through galaxy merging
\citep[e.g.,][]{pen07,jah11}.

The great interest in the origin of the scaling relations is reflected
in the flood of observational studies, focusing on their cosmic
evolution
\citep[e.g.,][]{tre04,wal04,shi06,mcl06,pen06,woo06,sal07,tre07,woo08,jah09,ben10,dec10,mer10},
with the majority pointing to a scenario in which BH growth precedes
bulge assembly.

However, many high-redshift studies to date are based on monochromatic
Hubble Space Telescope (HST) imaging, determining only the luminosity
of the host spheroid and not its stellar mass. This is acceptable at
$z\sim0.5$ \citep[e.g.,][]{ben10}, where the stellar populations of
bulges are fairly well known and their luminosities can be passively
evolved to zero redshift with uncertainties smaller than other sources
of error. In contrast, at $z>1$, the stellar populations of bulges are
an uncharted territory, particularly for AGN hosts which are believed
to be connected with major mergers and may have undergone recent
episodes of star formation \citep[e.g.,][]{kau03, san04}.  The
uncertainty on the conversion from observed luminosity to equivalent
$z=0$ luminosity can be comparable to the evolutionary signal
\citep[e.g.,][]{pen06}.

An exception is the study by \citet{mer10} who estimate total stellar
masses (\mts) and AGN luminosities by fitting spectral-energy
distribution (SED) models to multi-band data from the rest-frame
ultraviolet to the rest-frame mid-infrared for a sample of 89
broad-line AGN (BLAGN) hosts at $1<z<2.2$. Estimating \mbh\ from broad
MgII emission, they find that black holes of a given mass reside in
less massive hosts at higher redshift with a modest evolutionary slope.
However, \citet{mer10}
are unable to distinguish between \mts\ and the stellar mass of the
central bulge component of the host (\mss).  Such a difference may be
important when studying the evolution of the scaling relations: there
are indications that the relations between \mbh~and {\it total
host-galaxy} luminosity \citep{ben10} and stellar mass \citep{jah09}
may not be evolving, or at least not as rapidly as the relations
between \mbh\ and {\it spheroid} properties.  

In this paper, we study the cosmic evolution of the \mbh-\mss~and
\mbh-\mts~relations for a sample of 11 BLAGNs ($1<z<2$;
lookback time: 8-10 Gyrs) selected from
the Great Observatories Origins Deep Survey (GOODS) fields,
taking into account selection effects. \mss\ and
\mts~are derived from the deep multi-filter HST images. 
\mbh\ is estimated using the width of the broad MgII emission line,
measured from existing spectra and the 3000\AA~nuclear luminosity. 
We use a local comparison sample of Seyfert-1 galaxies \citep{ben11}
for which all relevant quantities were derived following
the same procedures adopted for the distant sample to minimize potential
systematic bias.
Our strategy allows us to address two major limitations of previous studies:
eliminate uncertainties due to luminosity evolution and determine the
evolution of the spheroidal component of the host.

Throughout the paper, we assume a Hubble constant of $H_0$ =
70\,km\,s$^{-1}$\,Mpc$^{-1}$, $\Omega_{\Lambda}$ = 0.7 and
$\Omega_{\rm M}$ = 0.3.
Note that all magnitudes are AB.

\section{DATA ANALYSIS}
\subsection{Sample Selection}
\label{sec:sample}
The high-redshift sample consists of AGNs in GOODS-N \citep{tre09} and
GOODS-S \citep{tro08,sil10}, selected based on their X-ray
emission using the Chandra Deep Field North and South (CDF-N/CDF-S)
survey and spectroscopically confirmed to be BLAGNs. 
We select all 11 objects within $1<z<2$ for which archival Very Large Telescope (VLT) and Keck
spectra covering the broad MgII line exist (Table~\ref{sample}).

\begin{figure*}
\begin{center}
\includegraphics[scale=0.265]{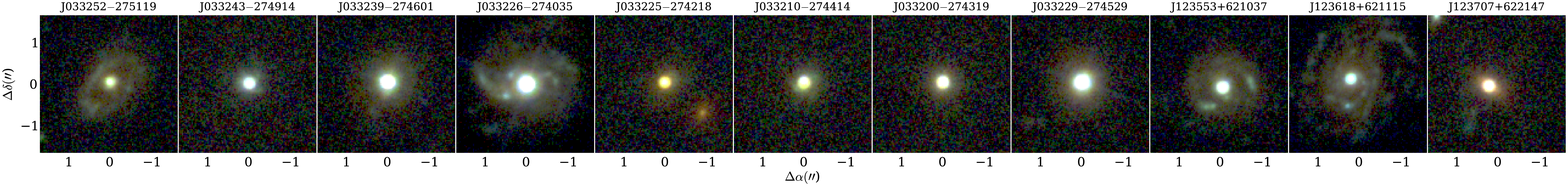}
\end{center}
\caption{
Deep HST/ACS color images (B, V, i, z), 3\arcsec$\times$3\arcsec. }
\label{sbp}
\end{figure*}

By design, all objects have deep HST/Advanced Camera for Surveys (ACS)
images in four different broad-band filters (B=F435W, V=F606W,
$i$=F775W, $z$=F850LP) \citep{gia04}.  Color images are shown in
Figure~\ref{sbp}.  The total exposure times range
between 5,000 and 25,000 sec, depending on the filter and the image
region.  The reduced data are taken from the v2.0 data
release.\footnote{http://archive.stsci.edu/pub/hlsp/goods/v2/}
The spatial resolution is approximately 0\farcs1 full-width-at-half-maximum (FWHM), which at $z=1.3$
(our average redshift) corresponds to 0.84 kpc; thus, our data have higher spatial resolution
than Sloan Digital Sky Survey (SDSS) images at $z=0.05$ (1\farcs4 = 1.37 kpc).
Overall, the AGN host galaxies look like typical ellipticals or
spirals, without any signs of merger activity.

Our local comparison sample consists of 25 Seyfert-1 galaxies
selected from SDSS ($0.02<z<0.1$;
\mbh$>10^{7}$M$_{\odot}$) for which all relevant quantities were derived following
the same procedures adopted for the distant sample to minimize potential
systematic bias \citep{ben11}.

\begin{deluxetable*}{lccccccccc}
\tabletypesize{\scriptsize}
\tablecolumns{11}
\tablewidth{0pc}
\tablecaption{Sample Properties, BH Masses, and Stellar Masses}
\tablehead{
\colhead{ID} & \colhead{R.A.} & \colhead{Decl.} & \colhead{$z$} & \colhead{Ref.} & \colhead{${\rm FWHM}_{\rm MgII}$} & \colhead{$\lambda L_{\rm 3000}$} & \colhead{$M_{\rm BH}$} & \colhead{$M_{\rm sph, \star}$} & \colhead{$M_{\rm disk, \star}$}\\
& (J2000) & (J2000) & & & (km\,s$^{-1}$) & (10$^{44}$\,erg\,s$^{-1}$)\\
\colhead{(1)} & \colhead{(2)} & \colhead{(3)} & \colhead{(4)} & \colhead{(5)} & \colhead{(6)} & \colhead{(7)} & \colhead{(8)} & \colhead{(9)} & \colhead{(10)}}
\startdata 
J033252$-$275119  &  ID.88  &  ID.8  &  1.227  &  S10/S04  &  16208  &  0.43  &   9.01  &   9.83  &    10.50 \\
J033243$-$274914  &  ID.24  &  ID.2  &  1.900  &  T09/S04  &  16381  &  1.77  &   9.31  &  10.64  &  \nodata  \\
J033239$-$274601  &  ID.09  &  ID.8  &  1.220  &  T09/S04  &   4344  &  5.38  &   8.39  &  10.54  &  \nodata  \\
J033226$-$274035  &  ID.50  &  ID.5  &  1.031  &  S10/S04  &   2430  &  9.51  &   8.00  &   9.53  &    10.75  \\
J033225$-$274218  &  ID.17  &  ID.8  &  1.617  &  S10/S04  &   4744  &  1.64  &   8.22  &  10.61  &  \nodata  \\
J033210$-$274414  &  ID.91  &  ID.9  &  1.615  &  S10/S04  &   5852  &  2.02  &   8.45  &  10.45  &  \nodata  \\
J033200$-$274319  &  ID.36  &  ID.7  &  1.037  &  S10/L05  &   3602  &  1.08  &   7.90  &   9.62  &  \nodata  \\
J033229$-$274529  &  ID.98  &  ID.9  &  1.218  &  S10/M05  &   5308  &  4.33  &   8.52  &  10.71  &  \nodata  \\
  J123553+621037  &  ID.13  &  ID.3  &  1.371  &  T08/W04  &   5441  &  2.32  &   8.41  &   9.99  &    10.84  \\
  J123618+621115  &  ID.58  &  ID.0  &  1.021  &  T08/W04  &   6988  &  1.19  &   8.49  &   9.29  &    10.95  \\
  J123707+622147  &  ID.46  &  ID.9  &  1.450  &  T08/W04  &  10654  &  1.57  &   8.91  &  10.74  &  \nodata  
\enddata				 		  
\tablecomments{ 
Column 1: target ID (based on R.A. and Decl.). 
Column 2: Right Ascension. 				  
Column 3: Declination.		  
Column 4: redshift (taken from Team Keck Redshift Survey (TKRS) \citep{wir04} for GOODS-N).
Column 5: reference for catalog from which objects were selected/
reference for origin of spectra. S10=\citet{sil10},
T09=\citet{tre09}, T08=\citet{tro08}; S04=\citet{szo04}, L05=\citet{lef05}, M05=\citet{mig05}, W04=\citet{wir04}.
Column 6: FWHM of broad MgII.			  
Column 7: rest-frame luminosity at 3000\AA~(fiducial error 0.1 dex).
Column 8: log $M_{\rm BH}/M_{\sun}$ (uncertainty: 0.5 dex).
Column 9: stellar spheroid mass log $M_{\rm sph, \star}/M_{\sun}$  (Chabrier IMF; uncertainty 0.3 dex for ellipticals, 0.4 dex for spirals). 					  
Column 10: stellar disk mass log $M_{\rm disk, \star}/M_{\sun}$  (Chabrier IMF; uncertainty 0.3 dex).
}
\label{sample}
\end{deluxetable*}

\subsection{Surface Photometry}
\label{sec:ima}
We perform two-dimensional surface photometry using the code ``Surface Photometry and
Structural Modeling of Imaging Data'' (SPASMOID) developed by one of
us (M.W.A.). The code allows a joint multi-band analysis of surface
brightness models, thus superceding the functionality of GALFIT
\citep{pen02}, and is described in detail in \citet{ben11}.  The
point-spread function (PSF) of the HST/ACS optics is modeled using the
closest bright star to a given object.  We impose a Gaussian prior on
the AGN colors with the mean given by the quasar composite spectrum
from \citet{van01} redshifted to the AGN redshift and with a $\sigma$
of 0.2~mag.  We model the host galaxy by either a single \citet{dev48}
profile or by a \citet{dev48} profile plus an exponential profile to
account for a disk.  Depending on the
images and residuals, we decide whether a given object is best fitted
by three components (PSF, spheroid, disk) or two components (PSF +
spheroid), as described by \citet{tre07,ben10,ben11}.
A disk component is evident in 4/11 objects. 
In all four cases, we can only clearly detect the bulge in the z band.

To probe the reliability of our AGN-host-galaxy decompositions
when using the blue restframe wavelengths
covered by the GOODS images, we tested the effect of bandpass shifting.
Given the host-galaxy morphology and level of activity 
of the sample studied here, a local sample
of Seyfert galaxies is a suitable comparison sample for this test.
(\citealt[]{scha10} also concluded that ``moderate luminosity AGN host galaxies
at $z\simeq2$ and $z\simeq0$ are remarkably similar''.)
We thus repeated the analysis of our local sample of 
AGN host galaxies \citep{ben11}, but now using only
ug SDSS photometry (instead of griz).
We are able to recover the
photometry of the bulge and point source to within 0.1 mags,
i.e.~smaller than our adopted systematic uncertainty, demonstrating that
bandpass shifting is not a concern within our level of precision.
Moreover, this is a conservative estimation, since the GOODS images at $z\simeq1.3$ 
not only cover wavelengths comparable to ug rest frame (F775W
and F850LP), but additionally also shorter wavelengths
(F606W and F435W), thus effectively providing more
information to disentangle point source and bulge. 
Furthermore, as already pointed out above, the GOODS images of $z\simeq1.3$ objects
have even higher resolution than SDSS images at $\simeq0.5$.

\subsection{Stellar Mass}
From the resulting magnitudes (Table~\ref{surface}), stellar masses
are estimated using a Bayesian stellar-mass estimation code
\citep{aug09} assuming a Chabrier initial mass function (IMF)
(Table~\ref{sample}). We impose conservative uncertainties of 0.3 dex on the
masses of the bulges (disks) for the bulge-dominated (disk-dominated) hosts.
The masses for the bulge components of the disk-dominated hosts are estimated by using the
$z$-band mass-to-light ratios of the bulge-dominated hosts in our sample that are at similar
redshifts; we therefore add in quadrature a 0.3 dex uncertainty, yielding a total stellar
mass uncertainty of 0.4 dex for these objects.
For two of our objects, \citet{scha10} report
stellar masses based upon template fits to the integrated light.  
Our results agree within the uncertainties
(assumed to be 0.2~dex for \citealt[]{scha10}).

\begin{deluxetable*}{lccccccccccccccccccccc}
\setlength{\tabcolsep}{0.0in} 
\tabletypesize{\scriptsize}
\tablecolumns{15}
\tablewidth{0pc}
\tablecaption{Results from Surface Photometry}
\tablehead{
\colhead{Object} & \multicolumn{4}{c}{PSF} & \multicolumn{4}{c}{Spheroid} & \multicolumn{4}{c}{Disk}\\
& B & V & i & z & B & V & i & z & B & V & i & z  \\
& (mag) & (mag) & (mag) & (mag) & (mag) & (mag) & (mag) & (mag) & (mag) & (mag) & (mag) & (mag) \\
\colhead{(1)} & \colhead{(2)} & \colhead{(3)}  & \colhead{(4)} & \colhead{(5)}  & \colhead{(6)}  & \colhead{(7)}  & \colhead{(8)} & \colhead{(9)} & \colhead{(10)} & \colhead{(11)} & \colhead{(12)}  & \colhead{(13)}}
\startdata
J033252$-$275119  &  25.03  &  23.84  &  23.40  &  23.47  &  \nodata  &  \nodata  &  \nodata  &  23.86  &    24.12  &    23.49 &    22.84  &    22.18 \\
J033243$-$274914  &  22.53  &  23.09  &  22.98  &  23.10  &    25.30  &    24.68  &    23.67  &  23.37  &  \nodata  &  \nodata &  \nodata  &  \nodata \\
J033239$-$274601  &  21.33  &  21.08  &  21.14  &  21.38  &    24.17  &    24.32  &    23.06  &  22.10  &  \nodata  &  \nodata &  \nodata  &  \nodata \\
J033226$-$274035  &  20.26  &  20.04  &  20.04  &  20.01  &  \nodata  &  \nodata  &  \nodata  &  23.27  &    22.22  &    21.87 &    21.22  &    20.63 \\
J033225$-$274218  &  25.38  &  23.87  &  22.84  &  22.50  &    25.08  &    24.61  &    23.84  &  23.08  &  \nodata  &  \nodata &  \nodata  &  \nodata \\
J033210$-$274414  &  24.24  &  23.19  &  22.61  &  22.74  &    24.13  &    24.10  &    23.55  &  22.96  &  \nodata  &  \nodata &  \nodata  &  \nodata \\
J033200$-$274319  &  22.81  &  22.42  &  22.45  &  22.47  &    43.90  &    25.33  &    24.00  &  23.04  &  \nodata  &  \nodata &  \nodata  &  \nodata \\
J033229$-$274529  &  21.24  &  21.31  &  21.42  &  21.76  &    23.80  &    23.32  &    22.49  &  21.66  &  \nodata  &  \nodata &  \nodata  &  \nodata \\
  J123553+621037  &  22.55  &  22.12  &  22.15  &  22.33  &  \nodata  &  \nodata  &  \nodata  &  24.04  &    23.62  &    23.20 &    22.45  &    21.83 \\
  J123618+621115  &  22.77  &  22.27  &  22.62  &  22.81  &  \nodata  &  \nodata  &  \nodata  &  23.87  &    23.15  &    22.37 &    21.40  &    20.79 \\
  J123707+622147  &  23.16  &  22.97  &  22.69  &  22.55  &    24.95  &    24.66  &    23.46  &  22.57  &  \nodata  &  \nodata &  \nodata  &  \nodata 
\enddata	   				
\tablecomments{    					 
Column 1: target ID.
Columns 2-5: extinction-corrected B, V, i, and z PSF magnitudes (uncertainty: 0.2 mag).
Columns 6-9: extinction-corrected B, V, i, and z spheroid magnitudes (uncertainty: 0.2 mag).
Columns 10-13: extinction-corrected B, V, i, and z disk magnitudes (uncertainty: 0.2 mag). 
}
\label{surface}
\end{deluxetable*}

\subsection{BH Mass}
\label{sec:spec}
Black hole masses are estimated via the empirically
calibrated photo-ionization method (``virial method'') \citep[e.g.,][]{wan99,ves06,mcg08},
by combining the FWHM of the broad
MgII\,$\lambda$2798\AA~emission line 
and the 3000\AA~AGN continuum luminosity \citep{mcg08}:
\begin{eqnarray*}
\log M_{\rm BH} = 6.767 + 2 \log \left(\frac{{\rm FWHM}_{\rm MgII}}{1000\,{\rm km\,s^{-1}}}\right)\\*
+ 0.47 \log \left(\frac{\lambda L_{3000}}{10^{44}\,{\rm erg\,s^{-1}}}\right)
\end{eqnarray*}
The AGN luminosity is derived from the PSF magnitudes in the filter
closest to rest frame 3000\AA, and extrapolated based on the assumed AGN SED of \citet{van01}
(\S\ref{sec:ima}; Table~\ref{sample}).

The nominal uncertainty of \mbh~using this method is 0.4 dex.
However, for some spectra, the low signal-to-noise (S/N)
makes the FWHM measurements uncertain by up to
$\sim$50\%, conservatively estimated.
Moreover, the spectra are not of sufficient quality
to remove the Fe emission which can result in 
overestimating the width 
of MgII by up to 0.03 dex (in FWHM, \citealt[]{mcg08}; see, however, \citealt[]{mer10}).
We therefore adopt an uncertainty of 0.5 dex.
Note that while we
used uniform priors for both the 3000\AA~luminosity 
and the black hole mass in our analysis,
employing more informed priors from the quasar luminosity function of
\citet{ric06}
or the black hole mass function of \citet{kel10} yield negligible changes to
our inference.

\subsection{\mbh$-M_{\star}$ Evolution}
\label{sec:evo}
Following and expanding on work by
\citet{tre07} and \citet{ben10},
we model the evolution of the offset of the \mbh-\mss~and 
\mbh-\mts~scaling relations by assuming a model of the form
\begin{eqnarray*}
{\rm log}M_{\rm BH} - 8 = \alpha\left[{\rm log}M_{*} - 10\right] + \beta
{\rm log}\left[1+z\right] + \gamma + \sigma
\end{eqnarray*}
where $\alpha$ is the slope of the relations at $z = 0$ and is assumed not
to evolve, $\gamma$ is the intercept of the relations at $z = 0$, and
$\sigma$
is the intrinsic scatter which is also assumed to be non-evolving. Here
$\beta$
describes the evolution of the scaling relation
(with $\beta = 0$ implying no evolution). We impose $\delta$-function priors of $\alpha = 1.09$
\citep{ben11} and $\gamma = -0.48$ for the $M_{\rm sph,*}$ relation and
$\alpha = 1.12$ \citep{har04} and $\gamma = -0.68$
for the $M_{\rm Host,*}$ relation; the priors on $\gamma$ were determined
by fitting to the local AGNs from \citet{ben11} while keeping the slope
fixed to the noted values. A normal distribution
prior is used for the intrinsic scatter with mean 0.4 and variance 0.01
and we employ a broad uniform prior for $\beta$. 
(Note that, strictly speaking, the variable $\sigma$ accounts for both
the intrinsic scatter in the relationship and the (much smaller) uncertainty on $\gamma$.)
We
use the $z = 1.0-1.2$ `elliptical' stellar mass function from
\citet{ilb10} to place priors on the stellar masses.
Furthermore, we include a prior on the
black hole masses that models our selection effects by
using a hard cutoff at the low mass end. This cutoff is
determined from the data and models lower limit of black hole masses
observable in each considered set of data.

The relation above is first fitted using the 11 galaxies in this sample.
The lower limit for the black hole masses assumed for the high
redshift objects is $10^{7.4} M_\odot$.
\citet{mer10} have independently tried
to infer the evolution of the \mbh-\mts~relation, but
their analysis is somewhat different than ours
(e.g.~IMFs, local comparison samples, definition of offset,
treatment of upper limits and selection effects).
We therefore also fit
the relation using the \citet{mer10} data (adjusted to a Chabrier IMF),
and we impose a limiting black hole mass of $10^{7.3}$ for these data.
The results of our inference are shown in Table \ref{TmodelFits}.
Given that the different fits to the \mbh-\mts~relation
(Merloni et al. data only, our data only, both combined)
result in the same $\beta$ within the uncertainties,
we adopt the one for the combined sample in the following.

\begin{deluxetable}{lcccc}
\setlength{\tabcolsep}{0.0in}
\tabletypesize{\scriptsize}
\tablecolumns{5}
\tablewidth{0pc}
\tablecaption{Evolving $M_{\rm BH}-M_{*}$ Scaling Relations}
\tablehead{\colhead{Model}  &  $\alpha$  &  $\beta$  &  $\gamma$  &
$\sigma$}
\startdata
$M_{\rm sph,*}^{a}$   &  1.09  &  $1.96\pm0.55$  &  -0.48  &  $0.36\pm0.1$ \\
$M_{\rm host,*}^{a}$  &  1.12  &  $1.68\pm0.53$  &  -0.68  &  $0.35\pm0.1$ \\
$M_{\rm host,*}^{b}$  &  1.12  &  $1.15\pm0.15$  &  -0.68  &  $0.16\pm0.06$ \\
$M_{\rm host,*}^{c}$  &  1.12  &  $1.11\pm0.16$  &  -0.68  &  $0.17\pm0.07$ 
\enddata
\tablecomments{
$^{a}$Fitted only using the 11 objects presented here.
$^{b}$Fitted using the objects presented here and the objects
from \citet{mer10}.
$^{c}$Fitted only using data from \citet{mer10}.
}
\label{TmodelFits}
\end{deluxetable}

\section{RESULTS AND DISCUSSION}
\label{sec:res}

4/11 AGNs are clearly hosted by late-type spiral galaxies, 
while the rest seem to be
spheroid dominated.  Keeping in mind the small-number statistics, the
fraction of disk-dominated host galaxies ($36\pm17$\%) is lower than
what has been found by \citet{scha10} ($80\pm10$\%) for 20 X-ray
selected AGNs at a comparable redshift ($1.5<z<3$) imaged by HST/Wide
Field Camera 3 (WFC3; F160W) with 1-2 orbits integration time.  One
difference is that our objects have higher X-ray luminosities
(0.5-8keV; $43.5<\log L_X<44.5$, mean=44.2, compared to $42<\log
L_X<44$, mean=43.1) which might
explain why we find a larger fraction of elliptical host galaxies.

Interestingly, none of the objects shows clear signs of interactions
or merger activity, while at redshifts of $z=0.4-0.6$, 32$\pm$9\%
of Seyfert-1s are hosted by interacting/merging galaxies
\citep{ben10}. However, we cannot exclude that some
of these low surface-brightness features might have been missed
\citep[see, e.g.][]{ben08}.  \citet{scha10} also do not report
interactions/mergers but their images are significantly shallower than
ours.  Star-forming galaxies at a redshift of $z \sim
2$, on the other hand, show a 33$\pm$6\% fraction of interacting or merging systems
\citep{foe09}.
\citet{scha10} interpret their high fraction of spiral galaxies as a
sign that secular evolution may play a non-negligible role in growing
spheroids and black holes. Our findings, including the lack of
merger activity, are consistent with such a scenario.

\begin{figure*}[ht!]
\begin{center}
\includegraphics[scale=0.4]{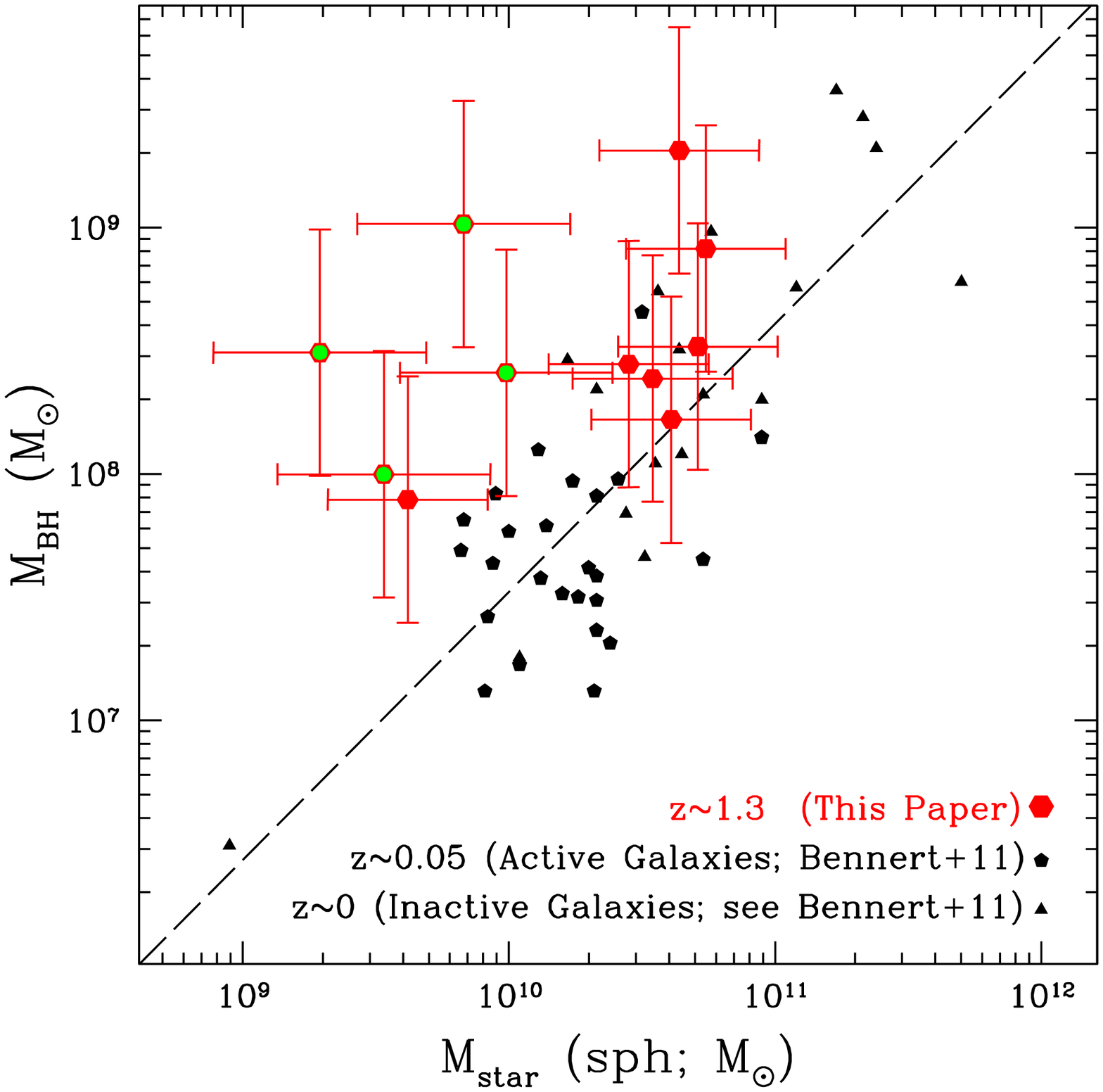}
\includegraphics[scale=0.4]{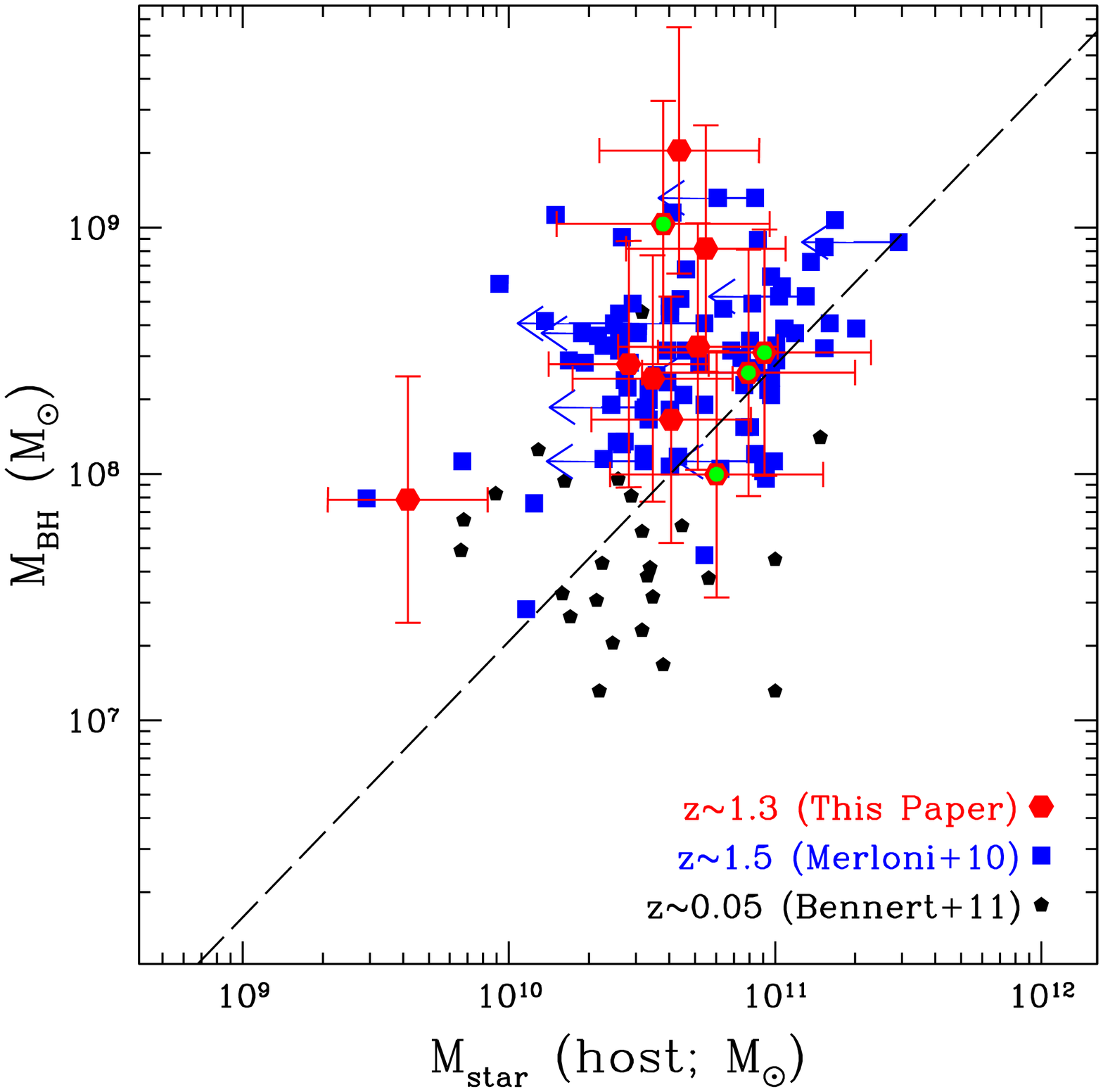}
\end{center}
\caption{{\bf Left panel:} \mbh-\mss~relation for our sample (red
pentagons; green circles: if fitted by spheroid plus disk), 
local BLAGNs \citep[black circles;][]{ben11} and
local inactive galaxies \citep[black triangles;][]{ben11}, with  $z=0$ relation
(see \S\ref{sec:evo}).  The errors for the local samples are omitted
for clarity (0.4 dex in \mbh, 0.25 dex in \mss).  {\bf Right panel:}
The same as in the left panel, for total host-galaxy stellar mass.
Here, we overplot the 89 BLAGNs from \citet{mer10} (blue filled
squares; 10 with upper limits indicated by arrows).}
\label{mbh_s}
\end{figure*}

Figure~\ref{mbh_s} (left) shows the \mbh-\mss~relation,
including a sample of 18 inactive galaxies 
and the local AGNs from \citet{ben11}.
In Figure~\ref{mbh_s} (right), we show the 
\mbh-\mts~relation, again including the local AGNs from \citet{ben11}
and additionally, the 89 AGNs from \citet{mer10} 
(10/89 with upper limits only;
subtracting 0.255 dex to convert their total
stellar masses from Salpeter to Chabrier IMF, \citealt[]{bru03}).
Note that for all comparison samples, 
\mbh~were estimated using the same recipe adopted here.

\begin{figure*}[ht!]
\begin{center}
\includegraphics[scale=0.4]{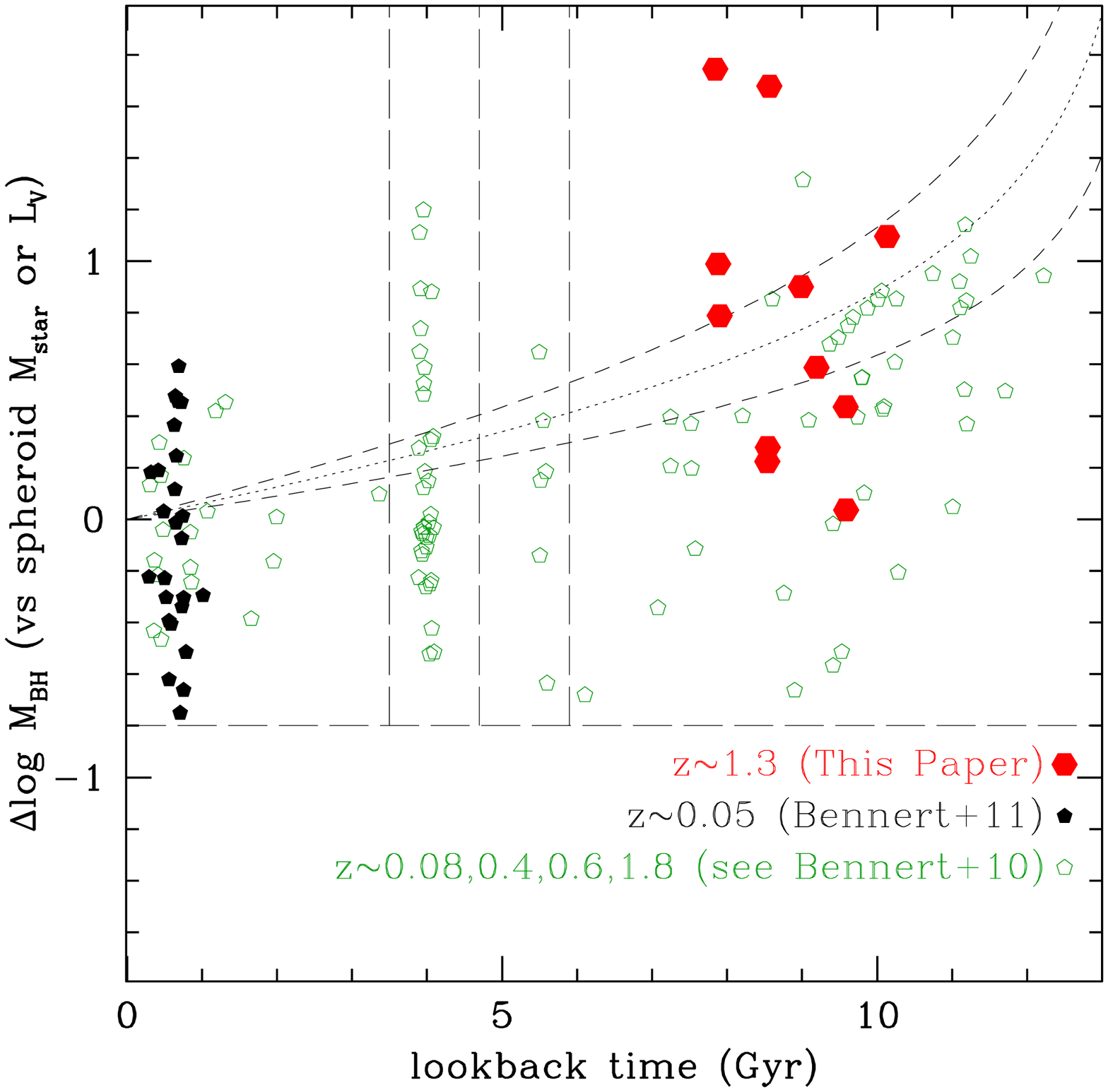}
\includegraphics[scale=0.4]{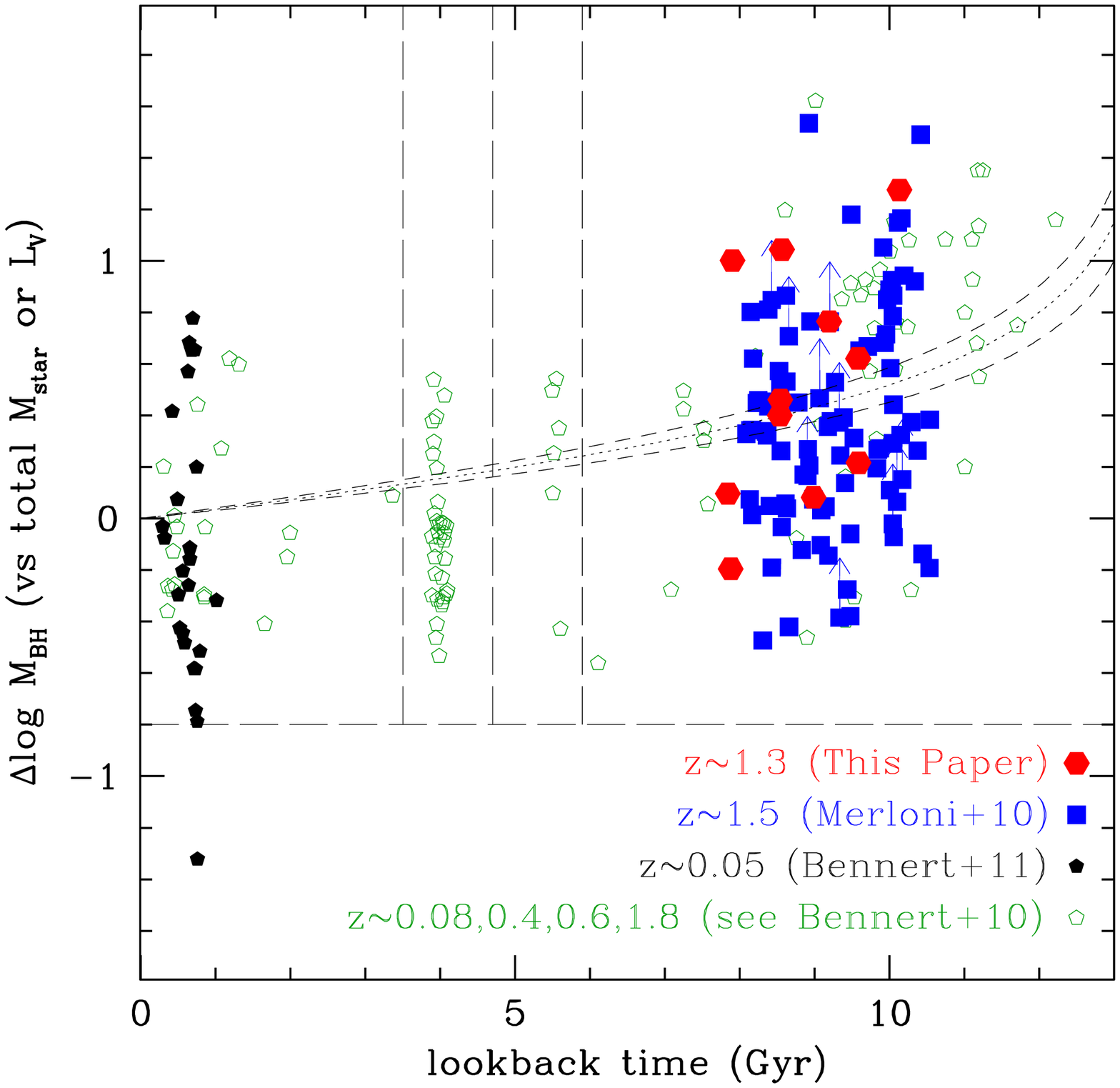}
\end{center}
\caption{ {\bf Left panel:} Offset in log \mbh~as a function of
constant \mss~(our objects: red filled pentagons) with respect to the
fiducial local relation of AGNs (black filled circles).  The offset in
log \mbh~as a function of constant stellar spheroid luminosity from
\citet{ben10} is overplotted (green open symbols), corresponding to
AGNs at different redshifts, (left to right: z$\simeq$0.08,
reverberation-mapped AGN from \citealt[]{ben10,ben09}; z$\simeq$0.4 from
\citealt[]{ben10,tre07} z$\simeq$0.6 from \citealt[]{ben10}, z$\simeq$1.8 from
\citealt[]{ben10,pen06}). The best linear fit derived here
is overplotted as dotted line
(\mbh/\mss~$\propto$ $(1+z)^{1.96\pm0.55}$;
dashed lines: 1$\sigma$ range).
{\bf Right panel:} The same as in the left panel as an offset in log \mbh~as a
function of constant total host galaxy mass (luminosity for
\citealt[]{ben10}).  The \citet{mer10} sample is overplotted (blue filled
squares).  The lines correspond to \mbh/\mts~$\propto$ $(1+z)^{1.15\pm0.15}$.}
\label{offset}
\end{figure*}

In Figure~\ref{offset}, we show the
offset in log \mbh~as a function of constant \mss~(left panel) and
\mts~(right panel) with respect to the $z=0$ relations
(see \S\ref{sec:evo}).  For
comparison, the offset in log \mbh~as a function of constant stellar
spheroid luminosity (left panel) and total luminosity (right panel)
from \citet{ben10} is overplotted.
Taking into account selection effects
(\S\ref{sec:evo}),
we find significant evolution in \mbh/\mss ($\propto(1+z)^{1.96\pm0.55}$), 
consistent with (but with larger uncertainties)
what we reported previously for the evolution
of the $M_{\rm BH} - L_{\rm sph}$ relation 
(\mbh/$L_{\rm sph}$$\propto(1+z)^{1.4\pm0.2}$; \citealt[]{ben10}). The agreement
between the stellar mass and luminosity evolution suggests that the
passive luminosity correction is appropriate, although modeling
luminosity evolution rather than stellar masses may increase the scatter.

For total stellar masses, including the \citet{mer10} data, the evolutionary
trend can be described as \mbh/\mts $\propto$ $(1+z)^{1.15\pm0.15}$,
in agreement with what has been found by \citet{mer10} within the uncertainties.
This evolution is slower than the one for spheroid masses ($\beta=1.96\pm0.55$)
in line with recent studies \citep{jah09,ben10}.
It indicates that the amount
by which at least some of the distant AGN host galaxies have to
grow their bulge component in order to fall on the local BH mass
scaling relations is contained within the galaxy itself.
It can thus be considered as another evidence that
secular evolution and/or minor mergers play a non-negligible role in growing
spheroids through a redistribution of stars
from disk to bulge. 
The deduced evolution is either in line
with or slightly faster than what has been predicted by theoretical
studies \citep[for a detailed comparison, see, e.g.][]{ben10,lam10}.

\section{CONCLUSIONS}
\label{sec:dis}
We determine spheroid and total stellar masses for the host galaxies
of 11 X-ray selected BLAGNs ($1<z<2$) in GOODS. In combination with
\mbh~estimated via the virial method from the broad MgII emission line
as measured from archival VLT and Keck spectra and the 3000\AA~nuclear
luminosity, we study the evolution of the \mbh-\mstar~scaling relation
out to a lookback time of 10 Gyrs.  Using a uniformly measured local comparison sample
and taking into account selection effects,
we find evolution of the
correlations consistent with BH growth preceding galaxy assembly,
confirming and extending the results of previous studies
\citep[e.g.,][]{mer10,dec10,ben10}.

Our results show that a significant fraction (4/11) of AGNs at z=1-2 are
hosted by spiral galaxies.
None of the galaxies show evidence for
recent major merger interaction, contrary to the general assumption
that BHs and spheroids grow predominantly through major mergers, a
scenario which might hold true only for the most luminous AGNs.
The evolution we find for the \mbh-total stellar
mass relation is slower than the one for spheroid stellar masses 
in line with recent studies \citep{jah09,ben10}. 
Combined, our results indicate that secular evolution and/or minor mergers play
a non-negligible role in growing both BHs and spheroids.

Our study demonstrates the feasibility of obtaining stellar masses of
AGN host galaxies out to lookback times of 10 Gyrs based on deep
multicolor HST photometry. This approach has the great advantage of
being independent of the luminosity evolution correction -- the
dominant source of systematic uncertainty in previous studies at
comparable redshifts \citep[e.g.][]{pen06,ben10}.  Furthermore, we can
distinguish between spheroid and total host galaxy mass, which is not
possible based on SED fitting \citep[e.g.,][]{mer10}.

Sample size is a major limitation of this work, allowing us to
constrain only average evolution and preventing us from investigating,
e.g, mass-dependent trends or correlations between evolution and
morphology. Follow-up of BLAGN hosts imaged by existing and
upcoming multicolor HST surveys (e.g. CANDLES) is needed to gather
larger samples and address theses remaining issues.

\acknowledgments

We thank Knud Jahnke, Andrea Merloni, and Kevin Schawinski for
discussions, and Brandon Kelly for providing the quasar BHMF.
We thank the anonymous referee for a careful reading of the manuscript
and valuable suggestions.
VNB, MWA, and TT acknowledge support by the NSF through
CAREER award NSF-0642621, and by the Packard Foundation through a
Packard Fellowship.  JHW acknowledges support by Basic Science
Research Program through the National Research Foundation of Korea
funded by the Ministry of Education, Science and Technology
(2010-0021558).  This research has made use of the public archive of
the SDSS and the NASA/IPAC Extragalactic Database (NED) which is
operated by the Jet Propulsion Laboratory, California Institute of
Technology, under contract with the National Aeronautics and Space
Administration.

\end{document}